# Pore and Ligament Size Control, Thermal Stability and Mechanical Properties of Nanoporous Single Crystals of Gold

*Maria Koifman Khristosov[a,b], Shiri Dishon[a], Imrit Noi[a], Alex Katsman[a], and Boaz Pokroy[a,b]\**

[a]Department of Materials Science and Engineering, Technion − Israel Institute of Technology, 32000 Haifa, Israel
[b]Russell Berrie Nanotechnology Institute, Technion − Israel Institute of Technology, 32000 Haifa, Israel

bpokroy@tx.technion.ac.il

**Abstract**

Nanoporous gold is widely used in research and nanotechnology because of its diverse properties, including high surface area and catalytic activity. The ligament size is usually considered as the only parameter controlling thermal stability and mechanical properties of nanoporous gold . Recently we developed a method for creating nanoporous single crystal gold particles using eutectic decomposition of Au-Ge, followed by selective etching of Ge. Here, we used this novel method to create nanoporous gold particles with controlled ligament sizes by changing the initial sample's relative concentrations of gold and germanium. When investigated over 1−4 h at 250−400°C the material was thermally stable up to 350°C, which is higher than the thermal stability of "classical" nanoporous gold with similar ligament sizes prepared by dealloying. Mechanical properties were examined utilizing nanoindentation of nanoporous gold before and after annealing. For smaller ligament sizes, hardness increases with annealing temperature up to 300°C and then strongly decreases. For larger ligament sizes, hardness decreases with increasing annealing temperature.Young's modulus was unchanged up to 300°C.

**Introduction**

Nanoporous gold is very promising in various fields of nanotechnology, such as in catalysis,[1-3] sensors,[4] actuators,[5] and electrodes for electrochemical uses[6] owing to its noble properties, high surface area, mechanical properties, and more. Its specific surface area is in the range of 3−10 $m^2g^{-1}$,[7, 8] and its yield strength is similar to or higher than that of bulk Au.[9] There are various ways to create nanoporous gold, the most common being by dealloying of an Au-Ag alloy in a nitric acid solution.[10] Other ways to obtain nanoporous gold are by growing gold in the presence of latex spheres,[11, 12] by growing nanopillar arrays,[13, 14] via nanoparticle-based gold,[15, 16] via nanofiber/nanotube mats,[17, 18] or by dealloying liquid metals.[19]

Thermal stability of polycrystalline nanoporous gold has been previously investigated in detail because of its importance in processes and applications requiring high temperatures, such as catalysis at elevated temperatures up to 200°C.[20, 21] Annealing at 300°C and above leads to damage of the nanoporous structure and coarsening of ligaments.[7] Thermally induced coarsening can even be utilized as a method for creating various ligament sizes of the nanoporous gold.[22, 23]

Mechanical properties of nanoporous gold were also widely investigated, especially after nanoporous gold was found not to fit the Gibson and Ashby scale relations for porous materials owing to the nanometric size of the ligaments.[24] Balk et al.[25] developed an updated scaling relation for the yield strength of nanoporous gold that took into account the nano-ligament size. Hamza et al.[26] showed that ligament size strongly affects nanoporous gold strength. Volkert et al.[27] investigated the strength of nanoporous gold under compression. However, the nanoporous gold used in most of those studies was prepared, unlike in the present study, by dealloying from an Au-Ag alloy.

We have previously shown that single crystals of nanoporous gold can be grown by eutectic decomposition of Au-Ge melts, followed by selective etching of the Ge phase.[28] We now describe a method for controlling the ligament and pore size of nanoporous gold by changing the relative concentrations of Au and Ge in the sample. We also present an investigation of the thermal stability and mechanical properties of particles made of nanoporous single crystals of gold as a function of ligament size.

**Experimental Section**

*Sample Preparation.* Silicon dioxide (thickness of 100 nm) was grown on (001) silicon wafers by thermal oxidation at 1100°C. Gold and germanium films (99.999% pure, Sigma-Aldrich) were successively evaporated onto the $SiO_2$ substrate in a PVD-4 thermal evaporator (Vinci Technologies) under a high vacuum of $10^{-6}$ Torr at room temperature, yielding a deposition rate of 3 Ås$^{-1}$. Gold and germanium films thickness was 120 and 80 nm, respectively (67at% Au, Sample 1), and 90 and 130 nm, respectively (48at% Au, Sample 2) (Table 1). Samples were thermally annealed at 550°C for 5 min in a MILA-5000 ULVAC-RIKO rapid thermal annealer in a forming gas environment ($N_2H_2$, 99.99%). Samples were wet-etched in two steps: (i) immersion in a solution of $NH_4OH:H_2O_2$ (1:25% vol) for 45 min followed by rinsing in deionized water; (ii) immersion in $HNO_3$ (70%) for 1 h, also followed by rinsing in deionized water. Thermal stability experiments were performed in a Linkam High Temperature Stage TS1000 in an ambient flow of Argon (99.999%).

*Sample Characterization.* Samples were imaged with a Zeiss Ultra Plus high-resolution scanning electron microscope (HRSEM). Cross sections were acquired with the FEI Helios NanoLab DualBeam G3 UC FIB. Nanoindentation was performed using a PI 85 PicoIndenter (Hysitron) with a Berkovich tip (radius ~200 nm) in load-control mode. Indentation was performed at a constant loading rate of 500 μNS$^{-1}$, with loads of 500−4000 μN.

**Results and Discussion**

**Pore and ligament size control.** Nanoporous single-crystal gold particles of two different pore sizes and ligament sizes were prepared by eutectic decomposition and selective etching of the Ge phase as described previously.[28] Both Sample 1 and Sample 2 were prepared as described in the Experimental Section (**Table 1**). Annealing of the samples above the eutectic temperature led to the melting of thin films into a hypereutectic (Ge excess) Au-Ge melt. Owing to the dewetting properties of the melt/substrate interface, droplets of this melt were formed on the SiO$_2$ surface. Upon cooling to room temperature, droplets of gold and germanium with eutectic microstructures were observed. After selective etching of the germanium, solid droplets consisting of spherical particles of porous gold were seen on high resolution scanning electron microscope (HRSEM) (**Figure 1**).

**Table 1**. Sample preparation and parameters.

|  | Au thickness [nm] | Ge thickness [nm] | Total at% of Au[a] | Relative density of gold in the eutectic phase[b] | Average ligament size [nm][c] |
|---|---|---|---|---|---|
| Sample 1 | 120 | 80 | 67 | 0.68 | ~70 |
| Sample 2 | 90 | 130 | 48 | 0.72 | ~220 |

[a]Total at% of Au was calculated from Au-Ge film thickness ratios. [b]Relative density of gold in the eutectic phase was calculated from HRSEM processing of an image taken from a cross section of the eutectic phase. [c]Average ligament size was calculated by measuring the narrowest point of each ligament.[29]

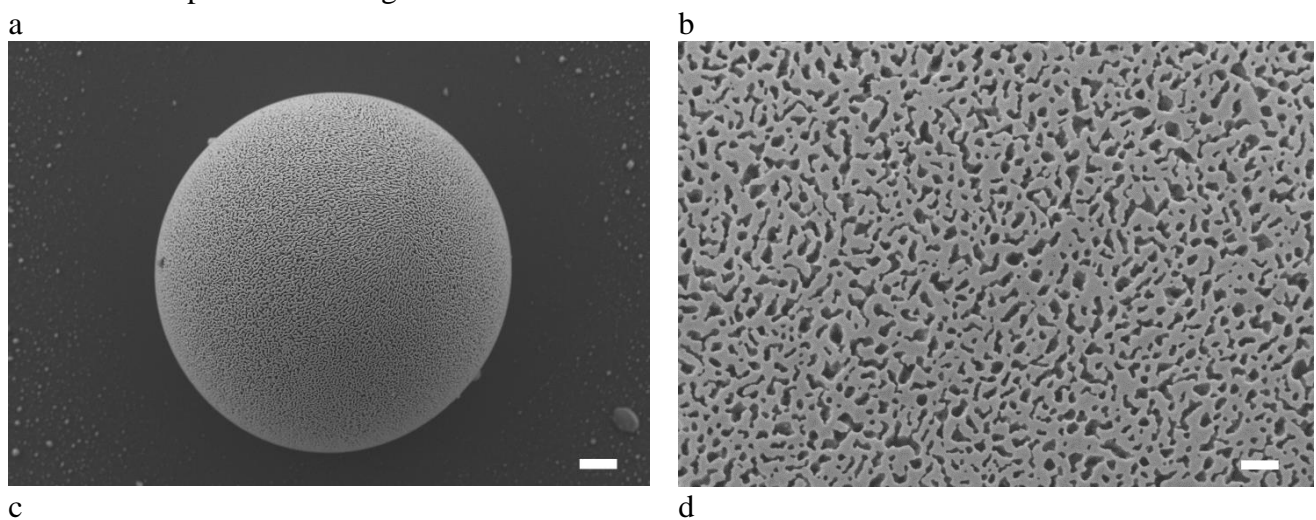

a  b

c  d

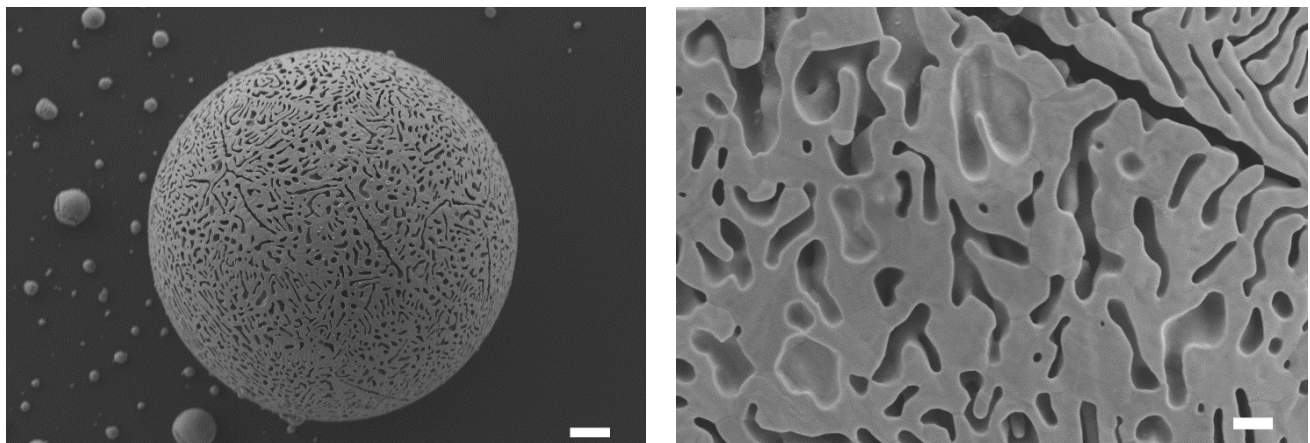

**Figure 1.** Single-crystal particles of nanoporous gold. (a) HRSEM image of a eutectic gold droplet after wet etching of germanium; top view, Sample 1. Scale bar, 1 μm. (b) HRSEM image of high magnification from (a). Scale bar, 300 nm. (c) HRSEM image of a eutectic gold droplet after wet etching, top view; Sample 2. Scale bar, 1 μm. (d) HRSEM image of high magnification from (c). Scale bar, 300 nm.

It can be seen from Figure 1 that by altering the relative concentrations of gold and germanium (via evaporation of different thicknesses during sample preparation), the ligament size of gold and germanium in the eutectic structure, and therefore the pore size, could easily be varied. For the higher (Sample 1) and the lower (Sample 2) Au concentrations, the ligament sizes were approximately 70 and 220 nm, respectively. The relative densities of gold in the eutectic phase of the two samples were almost the same, close to 0.7, enabling us to investigate this nanoporous gold as a function of ligament size at constant relative density.

In order to understand the influence of different Ge concentration on the ligament size, additional samples were evaporated with other Au-Ge ratios. Evaluation of these samples showed general tendency of higher ligament size for higher Ge concentration (**Figure S1**). A simple model taking into account heterogeneous nucleation of gold crystal from the eutectic melt on the liquid/Ge interface (Note S1) suggests that the larger amount of solid Ge phase before eutectic solidification results in smaller undercooling during eutectic structure growth. That is why the larger total (hypereutectic) Ge concentration results in the larger ligament size, $\lambda$, of the eutectic structure. According to the model, $\lambda \sim (X_{Ge} - X_{Ge}^{eut})^{2/3}$. Detailed and quantified description is given in **Note S1**.

**Thermal Stability.** Thermal stability experiments were performed at annealing temperatures in the range of 250−400°C and for different durations (1−4 h) (**Figure 2** and **Figure 3**).

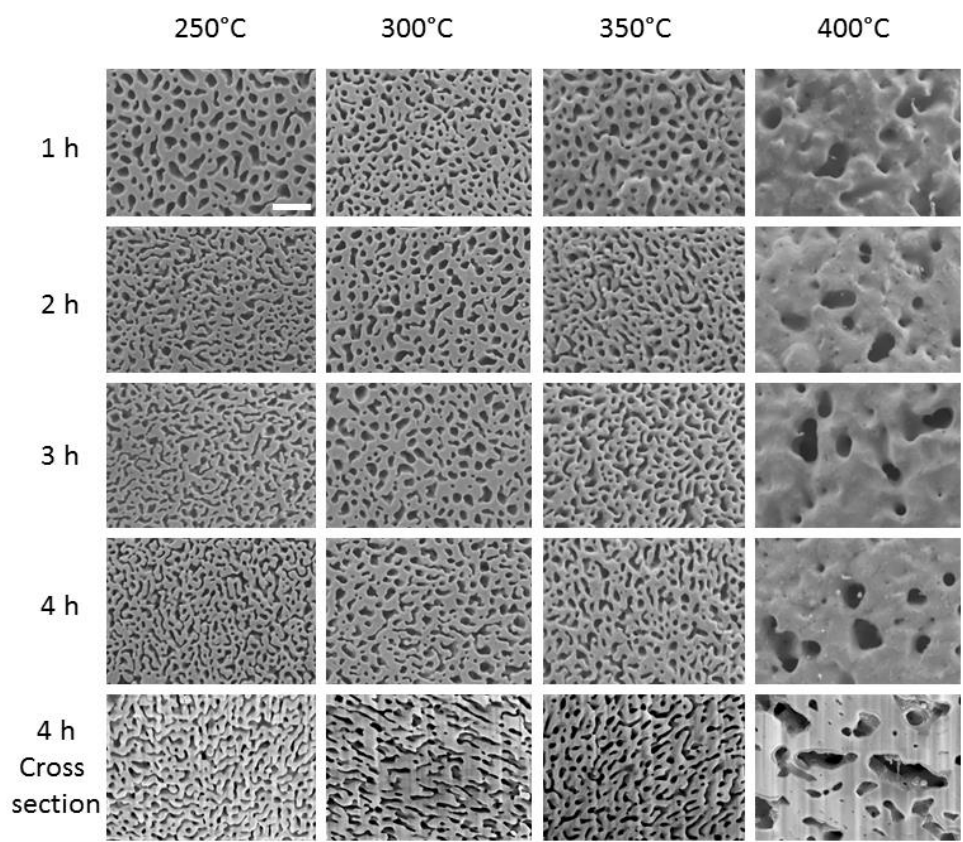

**Figure 2.** HRSEM images of nanoporous gold after thermal annealing experiments for Sample 1. Scale bar, 500 nm for all figures.

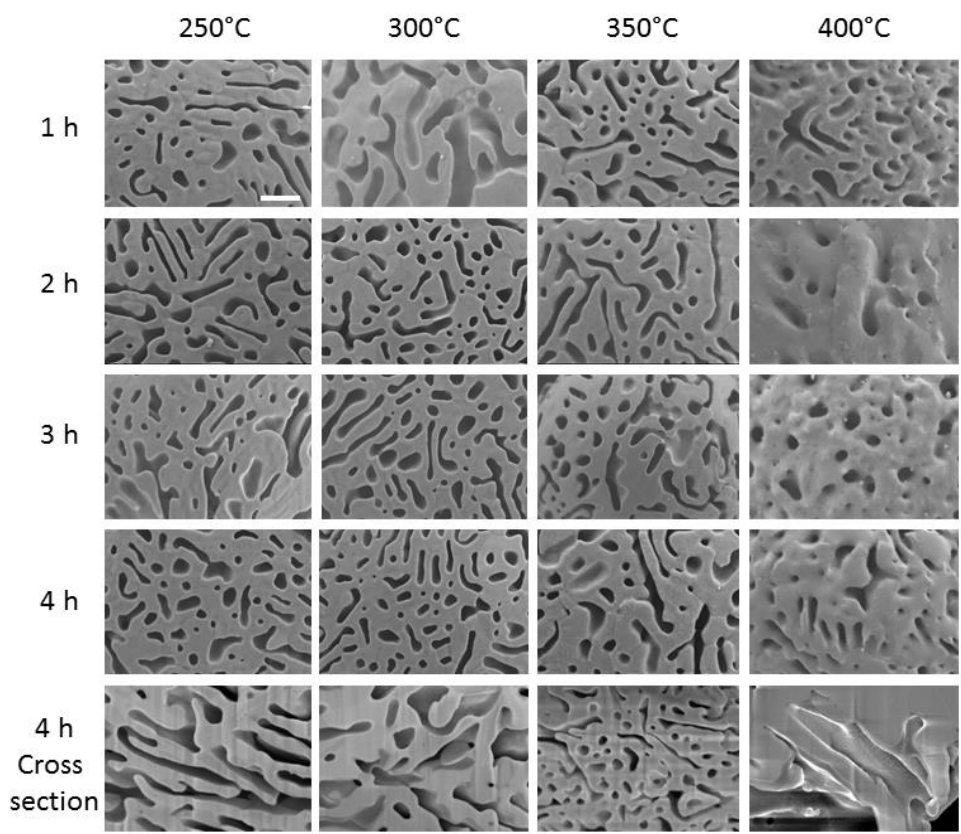

**Figure 3.** HRSEM images of nanoporous gold after thermal annealing experiments for Sample 2. Scale bar, 500 nm for all figures.

For both samples, as shown in Figure 2 and Figure 3, both the ligament and the pore nanostructures remained stable up to 350°C, with no detectable differences in microstructural morphology or pore size before and after annealing. At 350°C the pore surfaces underwent a slight change, becoming more rounded. In samples annealed for 1 h at 400°C, the porous nanostructure coarsened and the ligament size increased significantly. After annealing for 4 h at various temperatures we prepared cross sections of the droplets from Samples 1 and 2, in both cases using a focused ion beam (FIB) to compare the droplet's surface nanostructure with its internal nanostructure. In both samples the nanostructure remained stable and undamaged up to 350°C, and clear damage was visible only at 400°C.

These results demonstrated that the thermal stability of nanoporous gold prepared by eutectic decomposition, followed by selective etching of the Ge phase, is higher than that of nanoporous gold prepared by classical methods, such as dealloying of Au-Ag,[7] where after 2 h of annealing at 300°C the nanoporous structure is damaged. In nanoporous gold, coarsening of the ligaments is influenced by two diffusion processes: bulk diffusion and internal surface diffusion. Regarding bulk diffusion, we believe that the absence of grain boundaries (including small-angle sub-grain boundaries) in nanoporous gold prepared by this presented method results in relatively lower rates of self-diffusion, owing to the elimination of routes for the fast diffusion of grain-boundaries. This occurs because of growth of the whole single crystal inside the eutectic structure during eutectic solidification, as we showed in our previous work.[28] Dealloyed nanoporous gold, even if it is a large grain with a single orientation, still has low-angle grain boundaries that form during the dealloying process and serve as fast diffusion paths.

Regarding to the second process (internal surface diffusion), another reason for the relatively higher thermal stability of nanoporous gold prepared by this novel method could be that the pores formed during etching of germanium lamellas are "cylindrical-like" channels, mostly unconnected to each other. Therefore, diffusion along the channel surfaces cannot cause coarsening of the channels, but only a change in their shape. In dealloyed nanoporous gold, in contrast, the porous structure is interconnected and its evolution is controlled by the internal surface diffusion,[31, 32] which is much faster than diffusion across the bulk. Such a structure may therefore lose stability at rather low temperatures.

Evolution of this "cylindrical-like" channel porous structure, such as coarsening of closed channels or viscous sintering of porous material, requires diffusion across the bulk of material, whereas internal surface diffusion may lead only to morphological change (i.e., change in channel shape). This phenomenon was noted in particular in Sample 1. Therefore, evolution of the porous gold structure in this material is controlled by effective diffusion across the bulk with a typical distance of $l \sim 50$ nm between adjacent "cylindrical-like" voids.

The evolution of porous gold is controlled by an effective self-diffusion coefficient composed of bulk diffusion ($D_{bulk}$), grain-boundary (GB) diffusion ($D_{GB}$), and pipe diffusion through dislocations ($D_{disl}$):

$$D_{eff} = D_{bulk} + \left(\frac{\delta_1}{d}\right)D_{GB} + (\delta_2\sqrt{\rho})D_{disl} \tag{1}$$

where $\delta_1$ is the grain-boundary thickness, $d$ is the average grain size, $\delta_2$ is the dislocation core radius and $\rho$ is the dislocation density. The components of the diffusion coefficient are $D^0$ and $Q$ (pre-exponential coefficient and activation energy, respectively) in the gold. Each of the elements in **Equation 1** has been determined by different authors. For example, for self-diffusion in a gold single crystal, $Q_{bulk} = 1.73\pm0.04$ eV and $D_0^{bulk} = 1.1\div6.0$ cm$^2$s$^{-1}$;[33] for grain boundary self-diffusion in gold, $Q_{GB} = 0.88$ eV and $\delta_1 \cdot D_{GB}^0 = 3.1 \times 10^{-10}$ cm$^3$s$^{-1}$;[34] for dislocation-pipe diffusion, $Q_{disl} = 1.16\pm0.02$ eV and $\delta_2 \cdot D_{disl}^0 = 1.9\times10^{-10}$ cm$^3$s$^{-1}$.[35] From these values we can find the ratios of different terms in Equation 1 for different grain sizes and different dislocation densities (**Figure S2**). The grain boundary diffusion flux dominates at temperatures below 600÷650K for grain sizes with diameter of d ≤ 10 μm. In other words, diffusion-controlled processes in single crystals of gold should be much slower than in polycrystalline gold at temperatures T ≤ 650K.

Densification of porous material during heating can be considered as viscous sintering which involves macroscopic flow. Theory of viscous sintering is based on an assumption proposed by Frenkel,[36] that the energy dissipated in viscous flow is equal to the energy gained by the decrease in surface area during densification. For cylindrical pore structure one can apply analysis of Mackenzie and Shuttleworth[37] for viscous sintering of a body containing closed pores. To describe the sintering behavior of such a structure, it is necessary to choose a simple geometric form which retains the essential features of the real material. The model chosen (**Figure S3**) consists of an array of parallel porous cylindrical channels of a radius R spaced at a distance $l$ from each other in real incompressible material. The increase of density during annealing cannot be explained by volume diffusion of vacant lattice sites or surface migration of atoms, but must involve macroscopic flow. The driving force for this flow is surface tension, and an equation connecting the rate of shear strain with the shear stress defines the resistance to deformation. Using the approach of Mackenzie and Shuttleworth[37] the following expression for the time of full densification of a porous material can be obtained (**Note S2**):

$$\Delta t_s = (t_f - t_0) = \frac{2l_b^2 kT}{\gamma D_{eff} \Omega} \left(\frac{\pi - 2}{n}\right)^{1/2} \quad (2)$$

where $l_b$ is the characteristic size of microstructure, $\gamma$ is the surface tension, $\Omega$ is the atomic volume. Using the values for nanoporous gold: $\gamma = 1.28$ Jm$^{-2}$, $\Omega = 1.7 \cdot 10^{-29}$ m$^3$, $l_b = 1$ μm, $n = (1 - \rho_0)/\pi R^2 \rho_0$, where initial relative density $\rho_0 = 0.7$, and R = (25÷100) nm is the radius of the cylindrical channel, we may estimate the stability of porous structure against viscous sintering as the time required for full densification of the structure (**Figure S4**).

As can be seen, the nanoporous structure in a single gold crystal can remain stable during annealing at 600÷620K for about 4 h, whereas the presence of grain boundaries (for grain size d ≤ 1 μm) results in loss of stability (full sintering) at temperatures of 450÷470K.

In the case of nanoporous gold evolution, self-diffusion of the outer surface should also be taken into account. Activation energy of self-diffusion along the polycrystalline gold surface was found to be ~ 0.4 eV or even lower; see, for example ref. [38]. Evolution of near-surface voids was investigated by Kosinova et al.[39] Such evolution may change the morphology of the near-surface nanoporous structure, but it does not worsen stability of the whole nanoporous single crystal, at least at the annealing routes used in the present investigation.

**Mechanical Properties.** Mechanical properties of Samples 1 and 2 were investigated by nanoindentation, using a Berkovich tip in load-control mode. Nanoindentation was performed on top of the droplets using a constant loading rate of 500 μNs$^{-1}$, with loads of 500−4000 μN. Each droplet was subjected to only one loading test. Nanoindentation load-displacement curves obtained from nanoporous gold droplets of Samples 1 and 2 before annealing and after annealing for 4 h at 350°C are presented in **Figure 4**. For each sample the curves demonstrated marked overlapping in the loading section, indicating high reproducibility of the experiment.

(a)

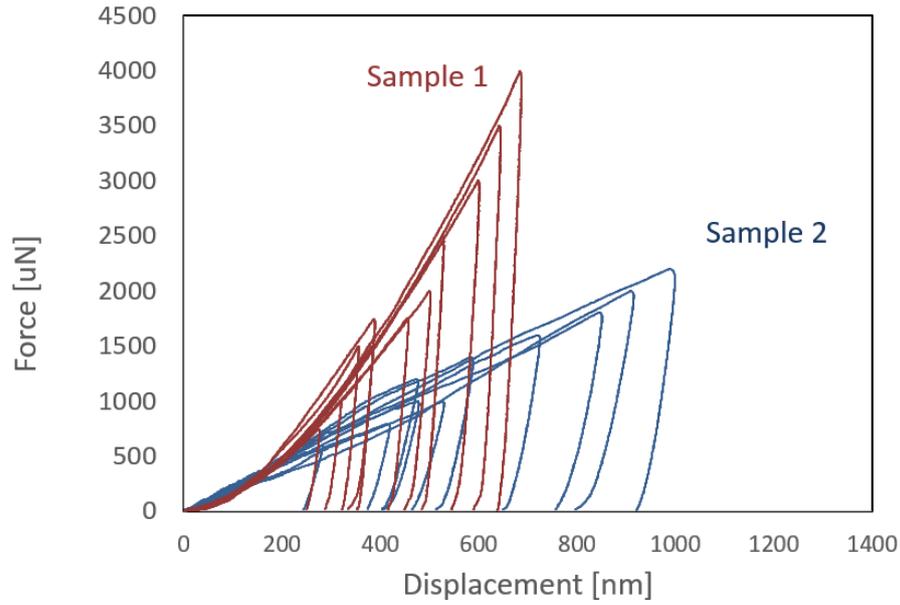

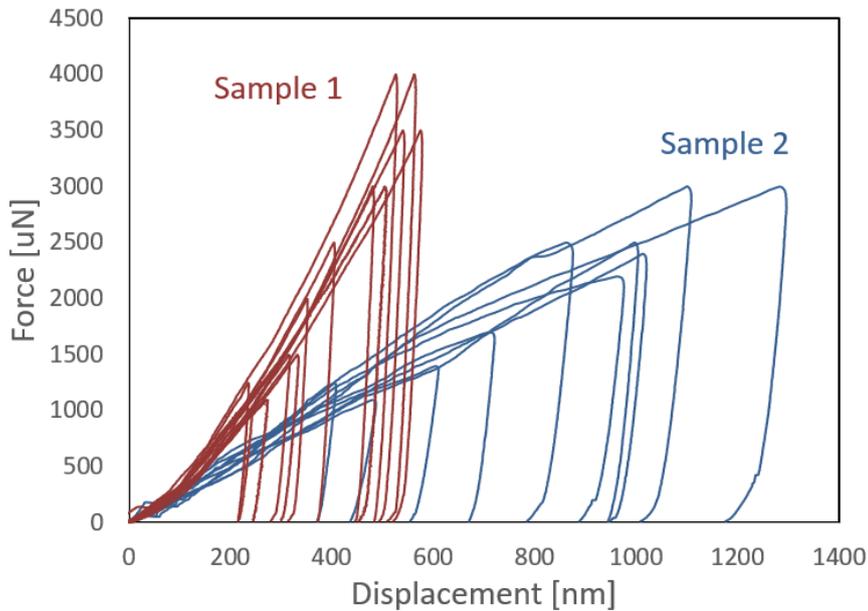

**Figure 4.** Nanoindentation experiments on nanoporous gold droplets from Samples 1 and 2, (a) before annealing and (b) after annealing for 4 h at 350°C.

Hardness (H) was calculated from load-displacement curves and is presented in **Figure 5(a-b)**. To prevent indentation-size effect, hardness was calculated from experiments in which the contact depth was larger than 350 nm, found to be the minimum value free of indentation-size effect.[40]

Reduced Young's modulus ($E_r$) was obtained from the load-displacement curves, and Young's modulus ($E_s$) of the sample was calculated from **Equation 3**:[40]

$$\frac{1}{E_r} = \frac{(1-\nu_i^2)}{E_i} + \frac{(1-\nu_s^2)}{E_s} \qquad (3)$$

where $\nu_i$ and $E_i$ are the Poisson's ratio and the Young's modulus of the nanoindenter (0.07 and 1140 GPa for a standard diamond indenter probe, respectively[40, 41]), and $\nu_s$ is the Poisson's ratio of the tested material (0.18 for nanoporous gold[42]) (Figure 5(c-d)).

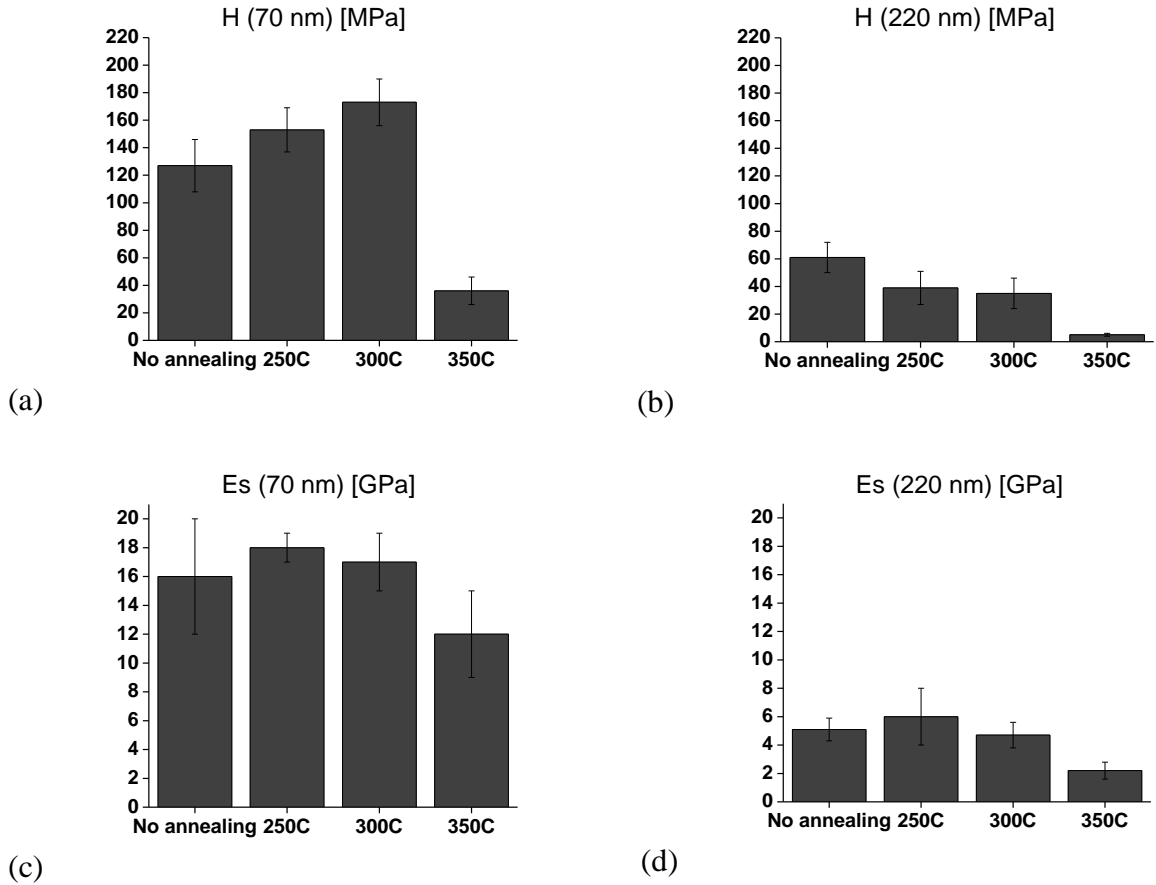

(a) (b) (c) (d)

**Figure 5.** Mechanical properties of Samples 1 and 2 (ligament size 70 nm and 220 nm, respectively), measured by nanoindentation. (a-b) Hardness (H); (c-d) Young's modulus ($E_s$). Young's modulus and hardness are clearly higher for nanoporous gold of smaller ligament size. $E_s$ is stable for both ligament sizes up to 300°C. Hardness behaviour differs for each sample, probably because of different mechanisms. For smaller ligament sizes, hardness increases with annealing temperature up to 300°C and then strongly decreases. For larger ligament sizes, hardness decreases with increasing annealing temperature.

The pre-annealing values of $E_s$ and H for nanoporous gold, obtained from nanoindentation, show good correlation with results from the literature and are in the reported ranges of similar materials.[25] The effects of ligament size on Young's modulus and hardness are clearly seen in Figure 5: as ligament size increases, $E_s$ and H decrease. This corresponds to results reported for nanoporous gold prepared by dealloying[43] and represents the known phenomenon of decreasing mechanical properties with increasing ligament size.[25]

Annealing up to 300°C has only a small effect on the Young's modulus in both samples, and at 350°C a small decrease is observed, reflecting the effect of annealing on the material's elasticity. In Sample 1, hardness values at 250°C and at 300°C were increased relative to the hardness measured before annealing. This could be related to hardening of the material owing to Ge residues (~0.5at.%) that remained in the sample after selective etching during sample preparation. Although most of the Ge canals—and therefore later the pores—are "open", a small amount of "trapped" Ge remained in closed pores after selective etching. In addition, there is up to 3at% solubility of Ge in Au, meaning that the gold phase also contains a small amount of Ge. In Sample 1 both the ligament size and the pore size are relatively small, and therefore the possibility of "trapped" Ge is high. In Sample 2, however, annealing leads to a decrease in hardness. In this sample the pores are significantly larger, and therefore probably contain no remaining "trapped" Ge in the droplets at 250°C and 300°C, but only the soluble Ge.

At 350°C a marked decrease in hardness was measured in both samples and could be explained as follows: remaining Ge creates eutectic melt with Au (eutectic temperature is 361°C), leading to a drastic decrease in both Young's modulus and hardness.

There are other speculated explanations. One possible reason for the change in mechanical properties implicates the movement of Ge traces from nanoporous gold "bulk" (inside the ligaments) to the surface as a result of annealing near the eutectic temperature. This could affect the surface energy of the nanoporous gold, leading in turn to changes in properties of the material. Yet another possible cause of the change could be surface dislocations. Annealing leads to surface diffusion of atoms on the surface of the ligaments, resulting in smoother pore surfaces, as seen in Figures 2 and 3. This in turn would result in fewer dislocations on the surface and lower hardness.

In the analysis of its mechanical properties, it is important to take account of the fact that this nanoporous gold is not a "bulk" homogeneous material, but is shaped like a droplet of approximately 10−20 μm in size. Its size effect should therefore be verified and eliminated, and the measured values of its mechanical properties should relate to the material itself and not be affected by the droplet's edges.

To address this issue, we evaluated the ratio of indentation depth to droplet size. The indentation depth was close to 500 nm and its height was approximately 15 μm, meaning that the indentation depth was only about 3% of the total droplet size. The nanoporous gold was compressed and densified during nanoindentation while the area outside the indentation remained undeformed, as in a similar experiment on "classical" nanoporous gold.[26] The above findings showed that both the size effect and the edges of the droplet can be neglected in this case. In addition, when we repeated these nanoindentation experiments on droplets of varying sizes, no influence of size on the mechanical properties was found. Therefore, although the particles subjected to nanoindentation were of finite size and not an "infinite" sample, we could assume that there were no size effects on hardness or elastic modulus.

**Conclusion**

The results of this study showed that nanoporous gold prepared by eutectic decomposition and selective Ge etching can be created with different and controllable ligament sizes. Our evaluations of the thermal stability of nanoporous gold showed superior stability as compared to "classical" nanoporous gold prepared by dealloying. This is attributed to the lower effective self-diffusion rates due to elimination of fast diffusion paths such as grain boundaries and interconnected internal surfaces, allowing the nanoporous single crystal structure to remain stable up to 350°C. The mechanical properties of the nanoporous single crystal gold measured by nanoindentation before and after annealing, showed that Young's modulus was unchanged up to 300°C, while its hardness varied differently for smaller and larger ligament sizes.

Advantages of our novel method include the ease of creating free-standing microparticles prepared by a relatively simple process with no need for additional fabrication steps, the simple control of ligament size, the higher thermal stability and good mechanical properties. All of these features further expand the functional possibilities of nanoporous gold and open the way to future new applications.

**Acknowledgment**
Thin films were fabricated at the Micro Nano Fabrication Unit (MNFU) at the Technion − Israel Institute of Technology, Haifa. We are grateful to Dr. Galit Atia, Dr. Alex Berner, and Mr. Michael Kalina for their help in preparing samples and operating the electron microscopes. We thank Prof. Menachem Bamberger for fruitful discussions. The research leading to these results received funding from the European Research Council under the European Union's Seventh Framework Program (FP/2007–2013)/ERC Grant Agreement no. 336077. M.K.K. is grateful for financial support by the Israeli Ministry of Science, Technology and Space.

**ToC figure**

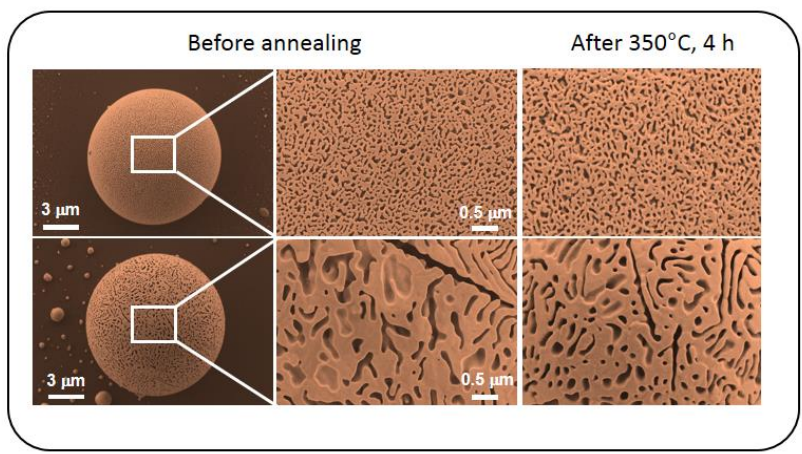